\documentstyle[twoside]{article}
\catcode`\@=11
\long\def\@makefntext#1{
\protect\noindent \hbox to 3.2pt {\hskip-.9pt  
$^{{\eightrm\@thefnmark}}$\hfil}#1\hfill}

\def\@makefnmark{\hbox to 0pt{$^{\@thefnmark}$\hss}}
\def\ps@myheadings{\let\@mkboth\@gobbletwo
\def\@oddhead{\hbox{}
\rightmark\hfil\eightrm\thepage}   
\def\@oddfoot{}\def\@evenhead{\eightrm\thepage\hfil
\leftmark\hbox{}}\def\@evenfoot{}
\def\sectionmark##1{}\def\subsectionmark##1{}}
\oddsidemargin=\evensidemargin
\newcounter{sectionc}\newcounter{subsectionc}\newcounter{subsubsectionc}
\renewcommand{\section}[1] {\vspace{12pt}\addtocounter{sectionc}{1} 
\setcounter{subsectionc}{0}\setcounter{subsubsectionc}{0}\noindent 
 {\tenbf\thesectionc. #1}\par\vspace{5pt}}
\renewcommand{\subsection}[1]
{\vspace{12pt}\addtocounter{subsectionc}{1} 
 \setcounter{subsubsectionc}{0}\noindent 
{\bf\thesectionc.\thesubsectionc. {\kern1pt \bfit #1}}\par\vspace{5pt}}
\renewcommand{\subsubsection}[1]
{\vspace{12pt}\addtocounter{subsubsectionc}{1}
 \noindent{\tenrm\thesectionc.\thesubsectionc.\thesubsubsectionc.
 {\kern1pt \tenit #1}}\par\vspace{5pt}}
\newcommand{\nonumsection}[1] {\vspace{12pt}\noindent{\tenbf #1}
 \par\vspace{5pt}}
\newcounter{appendixc}
\newcounter{subappendixc}[appendixc]
\newcounter{subsubappendixc}[subappendixc]
\renewcommand{\thesubappendixc}{\Alph{appendixc}.\arabic{subappendixc}}
\renewcommand{\thesubsubappendixc}
 {\Alph{appendixc}.\arabic{subappendixc}.\arabic{subsubappendixc}}
\renewcommand{\appendix}[1] {\vspace{12pt}
        \refstepcounter{appendixc}
        \setcounter{figure}{0}
        \setcounter{table}{0}
        \setcounter{lemma}{0}
        \setcounter{theorem}{0}
        \setcounter{corollary}{0}
        \setcounter{definition}{0}
        \setcounter{equation}{0}
        \renewcommand{\thefigure}{\Alph{appendixc}.\arabic{figure}}
        \renewcommand{\thetable}{\Alph{appendixc}.\arabic{table}}
        \renewcommand{\theappendixc}{\Alph{appendixc}}
        \renewcommand{\thelemma}{\Alph{appendixc}.\arabic{lemma}}
        \renewcommand{\thetheorem}{\Alph{appendixc}.\arabic{theorem}}
     \renewcommand{\thedefinition}{\Alph{appendixc}.\arabic{definition}}
       \renewcommand{\thecorollary}{\Alph{appendixc}.\arabic{corollary}}
        \renewcommand{\theequation}{\Alph{appendixc}.\arabic{equation}}
        \noindent{\tenbf Appendix \theappendixc #1}\par\vspace{5pt}}
\newcommand{\subappendix}[1] {\vspace{12pt}
        \refstepcounter{subappendixc}
        \noindent{\bf Appendix \thesubappendixc. {\kern1pt \bfit #1}}
 \par\vspace{5pt}}
\newcommand{\subsubappendix}[1] {\vspace{12pt}
        \refstepcounter{subsubappendixc}
       \noindent{\rm Appendix \thesubsubappendixc. {\kern1pt \tenit #1}}
 \par\vspace{5pt}}
\newcommand{\textlineskip}{\baselineskip=13pt}
\newcommand{\smalllineskip}{\baselineskip=10pt}
\def\eightcirc{
\begin{picture}(0,0)
\put(4.4,1.8){\circle{6.5}}
\end{picture}}
\def\eightcopyright{\eightcirc\kern2.7pt\hbox{\eightrm c}} 
\newcommand{\copyrightheading}[1]
 {\vspace*{-2.5cm}\smalllineskip{\flushleft
 {\footnotesize International Journal of Modern Physics B 11, 2207-2215 (1997)}\\
 {\footnotesize $\; $\, \ 
\ }\\
  }}

\newcommand{\publisher}[2]{{\begin{center}\footnotesize\smalllineskip 
 Received #1
 \end{center}
 }}
\def\abstracts#1#2#3{{
 \centering{\begin{minipage}{4.5in}\baselineskip=10pt\footnotesize
 \parindent=0pt #1\par 
 \parindent=15pt #2\par
 \parindent=15pt #3
 \end{minipage}}\par}}

\renewenvironment{thebibliography}[1]   
 {\frenchspacing
  \ninerm\baselineskip=11pt
  \begin{list}{\arabic{enumi}.}
 {\usecounter{enumi}\setlength{\parsep}{0pt}
  \setlength{\leftmargin 12.7pt}{\rightmargin 0pt} 
  \setlength{\itemsep}{0pt} \settowidth
 {\labelwidth}{#1.}\sloppy}}{\end{list}}
\newcounter{itemlistc}
\newcounter{romanlistc}
\newcounter{alphlistc}
\newcounter{arabiclistc}

\newcommand{\fcaption}[1]{
        \refstepcounter{figure}
        \setbox\@tempboxa = \hbox{\footnotesize Fig.~\thefigure. #1}
        \ifdim \wd\@tempboxa > 5in
           {\begin{center}
        \parbox{5in}{\footnotesize\smalllineskip Fig.~\thefigure. #1}
            \end{center}}
        \else
             {\begin{center}
             {\footnotesize Fig.~\thefigure. #1}
              \end{center}}
        \fi}
\newcommand{\tcaption}[1]{
        \refstepcounter{table}
        \setbox\@tempboxa = \hbox{\footnotesize Table~\thetable. #1}
        \ifdim \wd\@tempboxa > 5in
           {\begin{center}
        \parbox{5in}{\footnotesize\smalllineskip Table~\thetable. #1}
            \end{center}}
        \else
             {\begin{center}
             {\footnotesize Table~\thetable. #1}
              \end{center}}
        \fi}
\def\@citex[#1]#2{\if@filesw\immediate\write\@auxout
 {\string\citation{#2}}\fi
\def\@citea{}\@cite{\@for\@citeb:=#2\do
 {\@citea\def\@citea{,}\@ifundefined
 {b@\@citeb}{{\bf ?}\@warning
 {Citation `\@citeb' on page \thepage \space undefined}}
 {\csname b@\@citeb\endcsname}}}{#1}}
\newif\if@cghi
\def\cite{\@cghitrue\@ifnextchar [{\@tempswatrue
 \@citex}{\@tempswafalse\@citex[]}}
\def\citelow{\@cghifalse\@ifnextchar [{\@tempswatrue
 \@citex}{\@tempswafalse\@citex[]}}
\def\@cite#1#2{{$\null^{#1}$\if@tempswa\typeout
 {IJCGA warning: optional citation argument 
 ignored: `#2'} \fi}}

\def\pmb#1{\setbox0=\hbox{#1}
 \kern-.025em\copy0\kern-\wd0
 \kern.05em\copy0\kern-\wd0
 \kern-.025em\raise.0433em\box0}

\def\fnt#1#2{\footnotetext{\kern-.3em
 {$^{\mbox{\scriptsize #1}}$}{#2}}}
\def\fpage#1{\begingroup
\voffset=.3in
\thispagestyle{empty}\begin{table}[b]\centerline{\footnotesize #1}
 \end{table}\endgroup}
\def\runninghead#1#2{\pagestyle{myheadings}
\markboth{{\protect\footnotesize\it{\quad #1}}\hfill}
{\hfill{\protect\footnotesize\it{#2\quad}}}}
\font\tenrm=cmr10
\font\tenit=cmti10 
\font\tenbf=cmbx10
\font\bfit=cmbxti10 at 10pt
\font\ninerm=cmr9
\font\nineit=cmti9
\font\ninebf=cmbx9
\font\eightrm=cmr8

\def\qed{\hbox{${\vcenter{\vbox{   
   \hrule height 0.4pt\hbox{\vrule width 0.4pt height 6pt
   \kern5pt\vrule width 0.4pt}\hrule height 0.4pt}}}$}}

\def\bsc{{\sc a\kern-6.4pt\sc a\kern-6.4pt\sc a}} 
\def\bflatex{\bf L\kern-.30em\raise.3ex\hbox{\bsc}\kern-.14em 
T\kern-.1667em\lower.7ex\hbox{E}\kern-.125em X}

\begin{document}

\runninghead{D.~Mozyrsky, V.~Privman \& S.~P.~Hotaling}{Design of
gates for quantum computation: the NOT gate}

\normalsize\textlineskip
\thispagestyle{empty}
\setcounter{page}{1}

\copyrightheading{}  

\vspace*{0.88truein}

\fpage{1}
\centerline{\bf DESIGN OF GATES FOR QUANTUM COMPUTATION:}
\vspace*{0.035truein}
\centerline{\bf THE NOT GATE}
\vspace*{0.37truein}
\centerline{\footnotesize DIMA MOZYRSKY, \ VLADIMIR PRIVMAN}
\vspace*{0.018truein}
\centerline{\footnotesize\it Department of Physics, Clarkson University,
Potsdam, New York 13699-5820, USA} 
\vspace*{10pt}
\centerline{\normalsize and}
\vspace*{10pt}
\centerline{\footnotesize STEVEN P.~HOTALING}
\vspace*{0.015truein}
\centerline{\footnotesize\it Air Force Materiel Command, Rome
Laboratory/Photonics Division}
\baselineskip=10pt
\centerline{\footnotesize\it 25 Electronic Parkway, Rome,
New York 13441-4515, USA}
\vspace*{0.225truein}
\publisher{17 April 1997}{ }

\vspace*{0.21truein}
\abstracts{We offer an alternative to the
conventional network formulation of
quantum computing. We advance the {\it analog\/} approach to
quantum logic gate/circuit construction. As an
illustration, we consider
the {\it spatially extended\/} NOT gate as the first
step in the development
of this approach. We derive an explicit form of the interaction
Hamiltonian corresponding to this gate and
analyze its properties. We also
discuss general extensions to the case of certain
time-dependent interactions
which may be useful for practical realization of quantum
logic gates.}{}{}

\vspace*{1pt}\textlineskip 
\section{Introduction} 
\noindent
The fundamental physics of reversible quantum-mechanical
computation has received much 
attention recently.$^1$ Quantum computer is a hypothetical
quantum-coherent system that functions as 
a programmable calculational apparatus. Such a
computer will have to be drastically different from its classical 
counterparts. It will enable solution of certain problems$^1$ much
faster than the classical computer: the quantum interference property
yields$^1$ the fast-factoring (Shor's), as well as certain
other fast algorithms.
Recent theoretical results have included
identification of universal reversible 
two-bit gates$^2$ and advances$^3$ in error correction.
There have also been experiments$^4$ realizing a simple gate.

Nevertheless, the idea of construction of a 
macroscopic computer out of a large
number of quantum bits (qubits) is ellusive$^5$ at the present
stage of technology.  
The main obstacle is the sensitivity of coherent quantum evolution and 
interference
to undesirable external interactions such as noise or other failures in
operation.$^{1,5,6}$ Even though
a number of error correction schemes have
been proposed,$^3$ not all types of error
can be corrected. This particularly applies 
to the
{\it analog\/} nature of quantum
computers$^6$ which will be addressed below.

Quantum computers are naturally analog in their operation
because in order to use
the power of quantum interference, one has to allow any linear
combination of the basis qubit states. By analog errors we mean
those minor variations in the input and
output variables and in the system's 
dynamics
which cannot on their own be identified as erroneous in an analog device
because its operation involves
continuous values of variables (so that the
fluctuated values are as legal as the original ones). By noise errors we
mean those that result from single-event problems with device operation,
or from external influences, or from other failures in operation. In the
quantum case the latter errors also include the decoherence effects due
to influences of the environment.

Error-correction techniques can handle the noise errors but not
the analog errors. Indeed consider a state $\alpha |1\rangle
+\beta |0\rangle$ and a nearby state $\alpha
^{'}|1\rangle +\beta^{'}|0\rangle$, where $\alpha^{'}$ is close to
$\alpha$, while $\beta^{'}$ is close to $\beta$. 
Here $|1\rangle$ and
$|0\rangle$ denote the basis qubit or spin states
in the notation reminiscent
of the classical bit states 1 and 0.
Both linear-superposition states are equally legal
as input or output quantum states. Furthermore in the
conventional picture
of a quantum computer$^1$ which assumes a network
of a multitude of simple gate-units each
being controlled externally, the analog errors can proliferate and be
magnified in each step of the computation. 

In this work we therefore adopt a view typical of
the ``classical'' analog computer 
approach, of
designing the computer as a {\it single
unit\/} performing in one-shot a complex 
logical
task instead of a network of simple gates each performing
a simple ``universal-set''
logical function. In this case the computer as a whole
will still be subject to analog errors. However, these
will not be magnified by
proliferation of sub-steps each of which must
be exactly controlled. Indeed, 
quantum
(and more generally reversible) computation must be externally
timed: the time 
scale
of the operation of each gate is determined by the
interactions rather than by 
the relaxation
processes as in the ordinary computer. Furthermore, gate
interactions must be 
externally switched$^1$
on and off because the gates affect each other's operation.

In fact, we consider it likely that technological
advances might first allow design
and manufacturing of limited size units, based on several
tens of atomic two-level
systems, operating in a coherent fashion over sufficiently
large time interval to
function as parts of a larger classical (dissipative) computer
which will not
maintain quantum coherent operation over its macroscopic dimensions. We 
would
like these to function as
single analog units rather than being composed of many
gates.

While in principle in a reversible computational unit input and output
spins (qubits) need not be different, for larger units interacting with
the external world
it may be practically useful to consider input and output
separate (or at least not
identical). Indeed, the interactions that feed in the input 
need not
necessarily be identical to
those interactions/measurements that read off the
output.    

In light of these considerations we consider in this work a
{\it spatially extended\/} NOT gate
based on two spins: one input and one output.
Actually, we have to address a complicated set of problems: can 
multispin computational units be designed with short-range,
two-particle interactions? Can they
accomplish logical functions with interactions of the form familiar in
condensed matter or other experimental systems? These and similar
questions can only be answered by multispin-unit calculations
which will have to be numerical. Analytical results are limited
to the simplest gates such as NOT and XOR, the latter studied in
Ref.~7, and they provide only a partial picture.

This work is organized as
follows. In Section~2 we consider a simple, ``textbook'' 
example: 
the one-qubit NOT gate. It is
considered for illustration only and allows
us to introduce the
notation in a simple setting and exemplify some general ideas.
In Section~3 we consider the NOT gate with spatially separated
input and output qubits. The interaction Hamiltonian derived
for this gate, equation (21) below, 
establishes that it can be operated by the
internal interactions alone so that external-field effects can perhaps
be reserved for the clocking
of the internal interactions. Furthermore, it suggests the type of 
local internal interactions
to be used in more complicated systems where the computer as a whole
is treated as a many-body system with time-independent interactions.

The conventional formulation$^1$ of quantum computing involves
the external on and off switching of the interactions. In Section~4, we
show that this requirement can be relaxed and the
time dependence be given by other time-dependent
interactions (protocols) which are smoother than the on/off
shape. Section~4 also offers a summarizing discussion.

\section{The simple NOT gate}
\noindent
In this section we consider the NOT gate based on a two-state
system. Such a gate has been extensively studied in the 
literature,$^1$
so that our discussion is a review intended to set up the
notation and illustrate methods useful in more complicated situations.
We label by $\pmatrix{1\cr 0}$ and
$\pmatrix{0\cr 1}$ the two basis states.
The NOT gate corresponds to those
interactions which, over the time interval
$\Delta t$, accomplish the following changes:
\begin{equation}
\pmatrix{1\cr 0} \Longrightarrow
e^{i\alpha}\pmatrix{0\cr 1} \, ,
\end{equation}

\begin{equation}
\pmatrix{0\cr 1} \Longrightarrow
e^{i\beta}\pmatrix{1\cr 0} \, .
\end{equation}
The phases $\alpha$ and $\beta$ are
arbitrary. The unitary matrix $U$,
that corresponds to this evolution, is uniquely determined,
\begin{equation}
U=\pmatrix{0&e^{i\beta}\cr
              e^{i\alpha}&0} \, .
\end{equation}

The eigenvalues of $U$ are given by
\begin{equation}
u_1=e^{i(\alpha+\beta)/2} \qquad {\rm and}
\qquad u_2=-e^{i(\alpha+\beta)/2} \, ,
\end{equation}
while the eigenvectors, when
normalized and regarded as matrix columns,
yield the following transformation matrix $T$ which
can be used to diagonalize $U$:
\begin{equation}
T={1\over
\sqrt{2}}\pmatrix{e^{i\beta/2}&e^{i\beta/2}\cr
                  e^{i\alpha/2}&-e^{i\alpha/2}} \, .
\end{equation}
Thus, we have 
\begin{equation}
T^\dagger U T = \pmatrix{u_1&0\cr
                            0&u_2} \, .
\end{equation}
Here the dagger superscript denotes Hermitian conjugation.

We next use the general relation
\begin{equation}
U=e^{-iH\Delta t/\hbar}
\end{equation}
to identify the time-independent 
Hamiltonian in the diagonal representation.
Relations (4) yield the energy levels:
\begin{equation}
E_1=-{\hbar \over 2 \Delta t}(\alpha+ \beta)+{2\pi \hbar
\over \Delta t}N_1 \, , \;\;
 E_2=-{\hbar \over 2 \Delta t}(\alpha+ \beta)+{2\pi \hbar
\over \Delta
t}\left(N_2+{1\over 2}\right) \, ,
\end{equation}
where $N_1$ and $N_2$ are
arbitrary integers. The Hamiltonian is then obtained from the relation
\begin{equation}
H=T\pmatrix{E_1&0\cr
              0&E_2}T^\dagger \, 
\end{equation}
as a certain $2\times 2$ matrix. The latter is
conveniently represented is terms of
the unit matrix $\cal I$ and the conventional Pauli
matrices $\sigma_x$, $\sigma_y$, $\sigma_z$.
We get 
\begin{eqnarray} H&=&\left[-{\hbar \over 2 \Delta t}(\alpha+ \beta)+{\pi
\hbar\over\Delta t}\left(N_1+N_2+{1\over 2}\right)
\right]{\cal I}\vphantom{\Bigg]}\cr
&+&{\pi \hbar\over\Delta t}\left(N_1-N_2-{1\over 2}\right)\left[
\left(\cos{\alpha-\beta\over
2}\right)\sigma_x+\left(\sin{\alpha-\beta\over 2}\right)\sigma_y\right]
\vphantom{\Bigg]} \, .
\end{eqnarray}

To effect the gate operation, the interaction must be switched on
for the time interval $\Delta t$. The constant part of the interaction
energy (the part proportional to the unit matrix $\cal I$)
is essentially 
arbitrary; it only affects the average phase $\alpha+\beta \over 2$
of the transformation
(1)-(2). Thus this term can be omitted.

The nontrivial part of (10) depends on the integer $N=N_1-N_2$ which
is arbitrary, and on one arbitrary angular variable
\begin{equation}
\gamma = {\alpha- \beta\over 2} \, . 
\end{equation}
Thus we can use the Hamiltonian in the form
\begin{equation}
H={\pi \hbar\over\Delta t}\left(N-{1\over 2}\right)\left[
\left(\cos{\gamma}\right)\sigma_x
+\left(\sin{\gamma}\right)\sigma_y\right] \, .
\end{equation}
For a spin-$1\over 2$ two-state
system such an interaction can be obtained
by applying a magnetic field oriented
in the $XY$-plane at an angle $\gamma$ with the $X$-axis.
The strength of the field is inversely proportional to the desired
time interval $\Delta t$, and
various allowed field values are determined by
the choice of $N$.

We note that during application of the external field the {\it up\/}
and {\it down\/}
quantum states in (1)-(2) are {\it not\/} the
eigenstates of the Hamiltonian.
If the time interval $\Delta t$ is not short enough, the energy-level
splitting $|E_1-E_2|\propto
|N-{1\over 2}|$ can result in spontaneous emission
which is just one of the
undesirable effects destroying quantum coherence.
Generally, when implemented in a condensed
matter matrix for instance, the
two states of the qubit may lie within a spectrum of various
other energy levels.
In that case, in order to minimize the number of
spontaneous transition modes,
the best choice of the interaction strength would
correspond to minimizing
$|E_1-E_2|$, i.e., to $|N-{1\over 2}|={1\over 2}$.

\section{The spatially extended NOT gate}
\noindent
In this section we consider a spatially extended NOT gate consisting
of two spins: input and output. We will describe these spins
by four-state vectors and matrices labeled according to the following
self-explanatory convention: 
\begin{eqnarray} \pmatrix{a_1\cr a_2\cr a_3\cr a_4}
&=&a_1 |\uparrow\uparrow\rangle+
a_2 |\uparrow\downarrow\rangle+
a_3 |\downarrow\uparrow\rangle +a_4 |\downarrow\downarrow\rangle\cr
&=&a_1\pmatrix{1\cr 0}_I \otimes \pmatrix{1\cr 0}_O +
\; a_2\pmatrix{1\cr 0}_I \otimes \pmatrix{0\cr 1}_O
\vphantom{{\Bigg]}}\cr
&+&a_3 \pmatrix{0\cr 1}_I \otimes \pmatrix{1\cr 0}_O +
\; a_4 \pmatrix{0\cr 1}_I \otimes \pmatrix{0\cr 1}_O \, .
\vphantom{{\Big]\over 2}}
\end{eqnarray}
Here $I$ and $O$ denote {\it Input\/} and {\it Output}. In
what follows we 
will
omit the direct-product symbols $\otimes$ when multiplying expressions
with subscripts $I$ and $O$.

The desired transformation should take any state with $a_3=a_4=0$ into
a state with components 1 and 3 equal zero, i.e., {\it Input up\/}
yields {\it Output down}. Similarly, any state with $a_1=a_2=0$
should evolve into a state with components 2 and 4 equal zero,
corresponding to {\it Input down\/} giving {\it Output up}.
The general evolution operator must therefore be of the form
\begin{equation}
U=\pmatrix{0&0&U_{13}&U_{14}\cr
             U_{21}&U_{22}&0&0\cr
             0&0&U_{33}&U_{34}\cr
             U_{41}&U_{42}&0&0} \, ,
\end{equation}
which depends on 16 real parameters. However, one can show that the
requirement of unitarity, $U^\dagger U=1$,
imposes 8 conditions so that the number of
real parameters is reduced to 8. A lengthy but straightforward 
algebraic calculation then shows that the following parametrization
covers all such unitary matrices:
\begin{equation}
U=\pmatrix{0&0&e^{i\chi}\sin\Omega&e^{i\beta}\cos\Omega\cr
-e^{i(\alpha+\rho-\eta)}\sin\Upsilon&e^{i\rho}\cos\Upsilon&0&0\cr
0&0&e^{i\delta}\cos\Omega&-e^{i(\beta+\delta-\chi)}\sin\Omega\cr
e^{i\alpha}\cos\Upsilon&e^{i\eta}\sin\Upsilon&0&0} \, .
\end{equation}
Here all the angular variables are unrestricted although we could
limit $\Omega$ and $\Upsilon$ to the range $\left[0,{\pi \over
2}\right]$ without loss of generality.

In order to make the calculation analytically tractable, we will
restrict the number of free parameters to four by considering the case
\begin{equation}U=\pmatrix{0&0&0&e^{i\beta}\cr
             0&e^{i\rho}&0&0\cr
             0&0&e^{i\delta}&0\cr
             e^{i\alpha}&0&0&0} \, .
\end{equation}
This form has been favored for the following reasons. Firstly, the
structure of a single phase-factor in each column is similar to that of
the two-dimensional matrix encountered in Section~2. Secondly, the form
(16) contains Hermitian-$U$ cases ($\beta=-\alpha$, $\rho=0$ or $\pi$,
$\delta=0$ or $\pi$). Therefore, the eigenvalues, which are generally
on the unit circle for any unitary matrix, may be positioned more
symmetrically with respect to the real axis, as functions of the
parameters. These observations suggest
that an analytical calculation may be possible.

Indeed, the eigenvalues of $U$ turn out to be quite simple:
\begin{equation}
u_1=e^{i(\alpha+\beta)/2}\, ,  \;\;\;\;\;
u_2=-e^{i(\alpha+\beta)/2} \, ,  \;\;\;\;\;
u_3=e^{i\rho} \, ,  \;\;\;\;\;
u_4=e^{i\delta} \, .
\end{equation}
The diagonalizing matrix $T$ made up of the normalized
eigenvectors as columns is
\begin{equation}
T={1\over
\sqrt{2}}\pmatrix{e^{i\beta/2}&e^{i\beta/2}&0&0\cr
                  0&0&\sqrt{2}&0\cr
                  0&0&0&\sqrt{2}\cr
                  e^{i\alpha/2}&-e^{i\alpha/2}&0&0} \, .
\end{equation}

The next step in the calculation is to identify the energy levels.
We chose the notation such that the energies $E_{1,2}$ are identical
to (8). The other two energies are given by
\begin{equation} E_3=-{\hbar \over \Delta t}\rho +{2\pi \hbar
\over \Delta t}N_3 \, , \;\;\;\;\;
E_4=-{\hbar \over \Delta t}\delta +{2\pi \hbar
\over \Delta t}N_4 \, ,
\end{equation}
The Hamiltonian is then obtained
as in Section~2. It is convenient to
avoid cumbersome expressions by expressing it in terms of the energies;
the latter will be replaced by explicit expressions (8), (19) when
needed. The resulting $4\times 4$ matrix has been expressed in terms of
the direct products involving the unit matrices and the Pauli matrices
of the {\it Input\/} and {\it Output\/} two-state systems. This
calculation is
straightforward but rather lengthy. We only report the result:
\begin{eqnarray} H&=&{1\over 4}\left(2E_1+2E_2+E_3+E_4\right)
\;+\;{1\over
4}\left(E_3-E_4\right)\left(\sigma_{zI}-
\sigma_{zO}\right)\vphantom{\Bigg]}\cr
&+&{1\over
4}\left(2E_1+2E_2-E_3-E_4\right)\sigma_{zI}\sigma_{zO}
\vphantom{\Bigg]}\cr
&+&{1\over 4}\left(E_1-E_2\right)\left(\cos{\alpha-\beta\over 2}\right)
\left(\sigma_{xI}\sigma_{xO}-\sigma_{yI}\sigma_{yO}\right)
\vphantom{\Bigg]}\cr
&+&{1\over 4}\left(E_1-E_2\right)\left(\sin{\alpha-\beta\over 2}\right)
\left(\sigma_{xI}\sigma_{yO}+\sigma_{yI}\sigma_{xO}\right)
\, . \end{eqnarray}
As in Section~2, we note that the constant part of the Hamiltonian
can be changed independently of the other coupling constants and it can
be discarded. Recall that we can generally vary the integers
$N_{1,2,3,4}$ and the variables $\alpha$, $\beta$, $\rho$, $\delta$.
The ``constant'' part is in fact proportional to ${\cal I}_I \otimes
{\cal I}_O$. However, we avoid this cumbersome notation and present the
terms in the Hamiltonian in a more physically transparent form.

The Hamiltonian in (20) has also terms linear in the Pauli matrices (in
the spin components for spin systems). These correspond to interactions
with externally applied fields which in fact must be of opposite
direction for the {\it Input\/} and {\it Output\/} spins. We
try to avoid 
such interactions: hopefully, external
fields will only be used for ``clocking'' of the computation, i.e.,
for controlling the internal interactions of the {\it Input\/} and
{\it Output\/} two-state systems. Thus, we will assume
that $E_3=E_4$ so that 
there are no
terms linear is the spin components, in the Hamiltonian.

Among the remaining interaction terms, the term involving the
$z$-components in the product form $\sigma_{zI}\sigma_{zO}$ ($\equiv
\sigma_{zI}\otimes\sigma_{zO}$), has an arbitrary coefficient, say,
$-\cal E$. The terms of order two in the $x$ and $y$ components have
free parameters similar to those in (11)-(12) in Section~2.
The final expression is 
\begin{eqnarray}H&=&-{\cal E}\sigma_{zI}\sigma_{zO}
\vphantom{\bigg]}\cr
&&+
{\pi \hbar\over 2\Delta t}\left(N-{1\over 2}\right)\Big[
\left(\cos{\gamma}\right)\left(\sigma_{xI}
\sigma_{xO}-\sigma_{yI}\sigma_{yO}\right)\vphantom{\bigg]}\cr &&
 \hphantom{+{\pi \hbar\over 2\Delta t}\left(N{1\over 2}\right)\Big[l}
+\left(\sin{\gamma}\right)
\left(\sigma_{xI}\sigma_{yO}+\sigma_{yI}\sigma_{xO}\right)\Big]
\, .\end{eqnarray}
Here $N=N_1-N_2$ must be integer. In order to
minimize the spread of the energies
$E_1$ and $E_2$ we could choose $|N-{1\over 2}|={1\over 2}$ as in
Section~2. Recall that we already have $E_3=E_4$. Actually,
the energy levels of the Hamiltonian in the notation (21) are
\begin{equation} E_1=-{\cal E}+{\pi
\hbar\over \Delta t}\left(N-{1\over 2}\right)
\, , \;\;\;\; E_2=-{\cal E}-{\pi
\hbar\over \Delta t}\left(N-{1\over 2}\right)
\, , \;\;\;\; E_{3,4}={\cal E}\, .
\end{equation}
Thus further degeneracy (of three levels but not all four) can be
achieved by varying the parameters.

\section{Time-dependent interactions. Discussion}
\noindent
The form of the interactions in (21) is quite unusual as compared to
the traditional spin-spin interactions in condensed matter
models. The latter usually are based on the uniaxial (Ising)
interaction proportional to $\sigma_{z}
\sigma_z$, or the planar $XY$-model interaction proportional to
$\sigma_x\sigma_x +\sigma_y\sigma_y$, or the isotropic
(scalar-product)
Heisenberg interaction. The spin components here are those of two
different spins (not marked). The interaction (21) involves an
unusually high degree of anisotropy in the system. The $x$ and $y$
components are coupled in a tensor form which presumably will have to
be realized in a medium with well-defined directionality, possibly, a
crystal.
 
All the interaction Hamiltonians considered thus far were constant
for the duration of the gate operation. They must be externally
controlled. However, we note that the application of the interaction
need not be limited to the time-dependence which is an abrupt on/off
switching. Indeed, we can modify the
time dependence according to
\begin{equation}
H(t)=f(t)H \, ,
\end{equation}
where we use the same symbol $H$ for both the original
time-independent interaction Hamiltonian such as (21) and the new,
time-dependent one, $H(t)$. The latter involves the ``protocol''
function $f(t)$. The shape of this function, nonzero during the
operation of the
gate from time $t$ to time $t+\Delta t$, can be smooth.

For Hamiltonians involving externally applied fields, such as (12),
it may be important to have a constant plus an oscillatory components
(corresponding to constant and electromagnetic-wave magnetic fields,
for instance). However, 
the protocol function must satisfy
\begin{equation}
\int\limits_t^{t+\Delta t}f(t')\, dt'=\Delta t \, ,
\end{equation}
and therefore it cannot be purely oscillatory; it must have a
constant or other contribution to integrate to a nonzero value in
accordance with (24).

The possibility of the modification (23) follows from the fact that
the general relation
\begin{equation}
U=\left[e^{-i\int_t^{t+\Delta t}H(t')\, dt'/\hbar}
\right]_{\hbox{time-ordered}} \,
\end{equation}
does not actually require time ordering as long as the Hamiltonian
commutes with itself at different times. This condition is satisfied by
(23). Furthermore, if the Hamiltonian can be written as a sum of
commuting terms then each term can be multiplied by its own protocol
function. Interestingly, the Hamiltonian of the ``paramagnetic
resonance'' NOT gate, reviewed by DiVincenzo in Refs.~1, is not of
this form. It contains a constant part and an oscillatory part but
they do not commute. Note that the term proportional to $\cal E$
in (21) commutes with the rest of that
Hamiltonian. The terms proportional
to
$\cos \gamma$ and $\sin \gamma$ do not commute with each other though.
Rather, they anticommute, in (21), as such terms do in (12).

In summary, we have
derived expressions for the interaction Hamiltonians
appropriate for the NOT gate operation in two-state
systems. The expressions obtained will be useful in identifying
materials where there is hope of actually realizing such gates, in
writing down model Hamiltonians for more complicated, multispin
configurations, and in studying these gates as individual components,
for instance, with dissipation added.

\nonumsection{Acknowledgements}
\noindent
The work at Clarkson University has been supported in part by US Air
Force grants, contract numbers F30602-96-1-0276 and F30602-97-2-0089. 
The work at Rome Laboratory
has been supported by the AFOSR Exploratory Research Program and by
the Photonics in-house Research Program. This financial assistance 
is gratefully acknowledged.

\vfill
\vspace{12pt}\noindent{\tenbf
References}\vspace{1pt}\noindent\begin{thebibliography}{0}
\bibitem{1}
The following are general reviews: C. H.
Bennett, {\nineit Physics Today}, 
October 1995, p. 24; 
D. Deutsch, {\nineit Physics World}, June
1992, p. 57; D. P. DiVincenzo, {\nineit Science} {\ninebf 
270}, 255 (1995); 
A. Ekert and R. Jozsa, {\nineit Rev.
Mod. Phys.} {\ninebf 68}, 733 (1996); 
S. Lloyd, {\nineit Science} {\ninebf 261}, 
1563 (1993); B.
Schwarzschild, {\nineit Physics Today}, March 1996, p. 1.

\bibitem{2}
Recent literature includes A. Barenco, C. H. Bennett, R. Cleve, D. P. 
DiVincenzo,
N. Margolus, P. Shor, T. Sleator, J.
A. Smolin and H. Weinfurter, {\nineit Phys. Rev.}
{\ninebf A52}, 3457 (1995); 
D. P. DiVincenzo, {\nineit Phys. Rev.}
{\ninebf A51}, 1015 (1995); S. Lloyd, {\nineit Phys. Rev. Lett.} 
{\ninebf 75}, 346 (1995); 
A. Barenco, {\nineit Proc. R. Soc. Lond.} {\ninebf A449}, 679 (1995).

\bibitem{3}
For a recent review see D. P. DiVincenzo, {\nineit Topics in Quantum 
Computers\/} (preprint), as well as Refs.~1.

\bibitem{4}
C. Monroe, D. M. Meekhof, B. E. King, W. M. Itano and D. J. Wineland, 
{\nineit Phys. Rev. Lett.}
{\ninebf 75}, 4714 (1995); Q. Turchette, C. Hood, W. Lange, H. 
Mabushi and H. J. Kimble,
{\nineit Phys. Rev. Lett.}
{\ninebf 75}, 4710 (1995); see
also Refs.~1, and A. Steane, {\nineit The Ion 
Trap Quantum Information Processor\/} (preprint), as well as
N. A. Gershenfeld and I. L.
Chuang, {\nineit Science} {\ninebf 275}, 350 (1997).

\bibitem{5}
An instructive survey of these issues
and references to literature can be 
found in the article by R. Landauer, 
{\nineit Philos. Trans. R. Soc. London}
{\ninebf A353}, 367 (1995); see also 
S. Haroche and J.-M. Raimond, {\nineit Physics
Today}, August 1996, p. 51.

\bibitem{6}
A. Peres, {\nineit Phys. Rev.} {\ninebf A32},
3266 (1985); see also P. Benioff, {\nineit J. Stat. Phys.} 
{\ninebf 29}, 515 (1982).

\bibitem{7}
Details of a similar study for
the three-spin XOR gate have been reported 
by
D. Mozyrsky, V. Privman and S.
P. Hotaling, {\nineit Design of Gates for Quantum 
Computation: the three-spin XOR in
terms of two-spin interactions\/} (preprint). The 
XOR gate can be also realized in two-spin systems:
J. I. Cirac and P. Zoller, {\nineit Phys. Rev. Lett.}
{\ninebf 74}, 4091 (1995); I. L. Chuang and 
Y. Yamamoto, {\nineit The Persistent Qubit\/} (preprint); D.
Mozyrsky, V. Privman and 
M. Hillery, {\nineit Phys. Lett.} {\ninebf A226}, 253 (1997).
\end{thebibliography}
\end{document}